\documentclass[12pt]{article}
\usepackage{amssymb,amscd,amsmath,amsthm}
\usepackage{latexsym,amstext}
\usepackage{latexsym,amstext}
\usepackage{color}
\usepackage{graphicx}
\usepackage{epstopdf}
\usepackage{cite}
\usepackage[OT2,T1]{fontenc}
\DeclareSymbolFont{cyrletters}{OT2}{wncyr}{m}{n}
\DeclareMathSymbol{\Sha}{\mathalpha}{cyrletters}{"58}

\numberwithin{equation}{section}

\def\ds{\displaystyle}
\def\be{\begin{equation}}
\def\ee{\end{equation}}
\def\bea{\begin{eqnarray}}
\def\eea{\end{eqnarray}}
\def\bean{\begin{eqnarray*}}
\def\eean{\end{eqnarray*}}

\def\ds{\displaystyle}

\def\N{{\mathbb N}}
\def\Z{{\mathbb Z}}
\def\R{{\mathbb R}}
\def\C{{\mathbb C}}
\def\c{{\mathfrak C}}
\def\cw{{\widehat {\mathfrak C}}}

\def\d{{\mathfrak D}}

\def\H{{\mathcal H }}
\def\I{{\mathbb I }}

\begin{document}

\title{\bf Hermite functions and Fourier series}

\author{E. Celeghini$^{1,2}$, M. Gadella$^2$, M. A. del Olmo$^2$\\ \\
$^1$ Dipartimento di Fisica, Universit\`a di Firenze and\\  INFN-Sezione di Firenze\\
150019
Sesto Fiorentino, Firenze, Italy\\
$^2$ Departamento de F\'{\i}sica Te\'orica, At\'omica y Optica  and IMUVA, \\
Universidad de Va\-lladolid, 47011 Valladolid, Spain\\
}

\maketitle

\begin{abstract}Using
normalized Hermite functions, we construct bases in the space of square integrable functions on the unit circle  ($L^2(\mathcal C)$) and in   $l_2(\mathbb Z)$, which are related to each other by means of the Fourier transform and the discrete Fourier transform. These relations are unitary. 
The construction of orthonormal bases requires the use of the Gramm--Schmidt method. On both spaces, we have provided ladder operators with the same properties as the ladder operators for the one-dimensional quantum oscillator. These operators are linear combinations of some multiplication- and differentiation-like operators that, when applied to periodic functions, preserve periodicity. Finally, we have constructed riggings for both $L^2(\mathcal C)$ and  $l_2(\mathbb Z)$, so that all the mentioned operators are continuous.
\end{abstract}



\section{Introduction}\label{introduction}

Hermite functions have been an important tool in the development of elementary quantum mechanics as solutions of the quantum non-relativistic harmonic oscillator \cite{schiff}. From a mathematical point of view, Hermite functions serve as an orthonormal basis (complete orthonormal set) for the Hilbert space $L^2(\mathbb R)$. They are products of Hermite polynomials times and a Gaussian, so they are functions which are strongly localized near the origin \cite{RSI,szego}. 

The Fourier transform is a unitary operator on $L^2(\R)$. The Hermite functions are its eigenfunctions and allow a division of $L^2(\R)$ into four eigenspaces related to the cyclic group $C_4$. This division can be relevant in applications.

The one-dimensional Fourier transform and its inverse are automorphisms on the Hilbert space $L^2(\mathbb R)$, which preserve the Hilbert space norm, after the Plancherel theorem~\cite{RSII}. This result can be extended to some other spaces of interest in physics such as the space of infinitely differentiable functions converging to zero at the infinite faster than the inverse of a polynomial and the space of tempered distributions. In both cases the Fourier transform and the inverse Fourier transform are automorphisms, which are continuous with the standard topologies defined in both spaces \cite{RSII}.  

Loosely speaking, the Fourier series may be looked at as a particular case of a span of square integrable functions on a finite interval $[a,b]$ in terms of square integrable functions on this interval and extended beyond by periodicity.  One usually takes the interval $[0,2\pi]$ or $[-\pi,\pi]$. Then, one may look at the Fourier series as mappings from the space of $C^\infty$ functions on the unit circle to discrete  functions over the set of integers $\mathbb Z$. This mapping is invertible, which means that a sequence of complex numbers $\{a_n\}_{n\in\mathbb Z}$ with the property that $\sum_{n\in\mathbb Z}|a_n|^2<\infty$ uniquely fixes (almost elsewhere) a square integrable function on the unit circle \cite{BN}. These numbers are obtained using a discrete-time Fourier transform \cite{oppenheim} over the original function, so that the Fourier series and discrete-time Fourier transforms may be considered as operations inverse of each other. 

In the present article, we construct a complete sequence of periodic functions using the Hermite functions, which is a non-orthonormal basis on $L^2[-\pi,\pi] \equiv L^2(\mathcal C)$, where $\mathcal C$ is the unit circle. Then, after the Gram--Schmidt procedure we obtain an orthonormal basis formed by periodic functions. All functions on this orthonormal basis can be spanned into a Fourier series with coefficients obtained from the Hermite functions. Vice-versa, these coefficients are obtained via the discrete Fourier transform of the functions belonging to the orthonormal basis. 

We proceed to an equivalent construction on the space $l_2(\mathbb Z)$ of square summable complex sequences indexed by the set of integer numbers. Each term of the sequence is given by the value on an integer of a given normalized Hermite function. Fourier series and Fourier transform give the equivalence between both constructions, which induces a unitary mapping between both spaces.

We also construct multiplication and differentiation operators on a subspace of $L^2(\mathcal C)$ that includes our complete sequence of periodic functions in a way that these operators preserve periodicity. They play a similar role to that played by multiplication and derivation operators  in the description of the one-dimensional quantum harmonic oscillator. In particular, they provide ladder operators for the  complete sequence of periodic functions analogous to the creation and annihilation operators for the quantum oscillator. We may also construct a locally convex space of test functions, $\mathfrak S$, including the chosen complete sequence of periodic functions, such that these operators are continuous on $\mathfrak S$. This space, being dense in $L^2(\mathcal C)$ with continuous injection, defines a rigged Hilbert space $\mathfrak S \subset L^2(\mathcal C)\subset\mathfrak S^\times$, so that the operators can be continuously extended to the dual $\mathfrak S^\times$. A similar construction is also possible for $l_2(\mathbb Z)$. 

The rigging of Hilbert spaces is necessary for the continuity of  
the relevant operators that we introduce in this presentation and this includes the new  ladder operators. The formalism of rigged Hilbert spaces was introduced by Gelfand \cite{GEL}. Although this is not quite familiar to many theoretical physicists, it has acquired more and more importance in the field of mathematical physics \cite{BOH,RO,ANT,MEL,GG,GG1} and even in mathematics \cite{BAUM,BT,BBT,CHI,CHI1,FEICH,FEICH1}.

In a series of previous articles \cite{CGO16,CGO17,CGO18,CGO19,CGO20}, we have discussed the relations between discrete and continuous basis, algebras of operators, special functions and rigged Hilbert spaces (also called Gelfand triplets). We have shown that all these concepts properly convive in the framework of rigged Hilbert spaces and not on Hilbert spaces. The present paper fits also in part within the same context, where our special functions are now series constructed from Hermite functions.  In addition, it can be shown that the structure of rigged Hilbert spaces is suitable for the representation of our operators as an algebra of continuous operators on the same domain, while on Hilbert space these operators are unbounded. Thus, we create a representation with some formal and mathematical advantages derived from this continuity on a common domain.

The organization of this paper is as follows: In Section \ref{secondsection}, we study the behavior of the Hermite functions under the Fourier transform. The well-defined parity of the Hermite functions allows that the even/odd Hermite functions span the subspace of the even/odd functions of $L^2(\R)$. In a similar way,  since the Hermite functions are eigenvectors of the Fourier transform with eigenvalues the four roots of unity, $L^2(\R)$ can be split in four subspaces of functions,
 each one characterised by one eigenvalue of the Fourier transform as an operator on $L^2(\R)$.
In Section \ref{periodicfunctions}, we introduce a set of functions defined in the unit circle in terms of the Hermite functions. In Sections \ref{FTcircle} and \ref{discretefourier},
we present some results with the aim of given a unitary view that include different concepts such as  Fourier transform, Fourier series, discrete Fourier transform and Hermite functions. Later in Sections \ref{relevantoperators} and \ref{continuity}, we define ladder operators and those others related to them and the equipations of Hilbert spaces, or Gelfand triplets, or rigged Hilbert spaces on which these operators are well defined as continuous operators. We finish the present article with concluding remarks plus two Appendices.   In Appendix \ref{apendb}, we show two results which are important in the development of the material in Section \ref{discretefourier}. In Appendix \ref{apendc}, we construct orthonormal systems in $L^2(\mathcal C)$ and $l_2(\mathbb Z)$  via the Gramm--Schmidt process.


\section{The Hermite Functions and the Fourier Transform}\label{secondsection}

Let us consider the normalized Hermite functions in one-dimension, sometimes also called the Gauss--Hermite functions  \cite{NIST,EM,EM2015,E2015}. As is well known, they have the following~form
\begin{equation}\label{2.1}
\psi_n(x):=\frac{e^{-x^2/2}}{\sqrt{2^n\,n!\,\sqrt\pi}}
\;H_n(x)\,,
\end{equation}
where $H_n(x)$ are the Hermite polynomials, $x\in \R$ and $n \in \N$, with $\N$  the set of non-negative integers, $0,1,2,\dots$.    It is well-known that the Hermite functions  form an orthonormal basis in the Hilbert space of square integrable functions on the line, $L^2({\mathbb R})$. In other words, they verify the following orthonormality and completeness  relations
\begin{equation}\begin{array}{l}\label{2.2}
\displaystyle
\int_{-\infty}^\infty \psi_n(x)\psi_m(x)\,dx=\delta_{nm}\,,\\[0.4cm]
\displaystyle
\sum_{n=0}^\infty \psi_n(x)\psi_n(x')=\delta(x-x')\,.
\end{array}\end{equation}
 Observe that the completeness for the Hermite functions is given and labelled by the natural numbers, so that $n \in \N$, although the usual definition of the Parseval identity is a sum on the set of integers. This is not a contradiction, so both the set of natural numbers $\N$ (including the zero) and the set of integers $\Z$ are both infinite countable. One may associate the even Hermite functions, $\psi_{2n}$, to the positive integers and the odd Hermite functions, $\psi_{2n+1}$, to the set of negative integers. 

The Hermite functions   \eqref{2.1} are eigenfunctions of the Fourier Transform (FT)  in the following sense \cite{santhanam, CGO16}
\begin{equation}\label{2.3}
\text{FT}[\psi_n(x),x,y] \equiv \widetilde\psi_n(y) :=
\frac 1{\sqrt{2\pi}}  \int_{-\infty}^\infty e^{ixy}\,\psi_n(x)\,dx
= i^n \psi_n(y)\,,
\end{equation}
or more generally,
\begin{equation}\label{2.4}
\text{FT}[\psi _n(a x+b),x,y]=\frac{i^n}{|a|}\, e^{-iby/a}\,\psi _n (y/a)\,,\qquad a\neq 0\,,\; b \in \R\,,
\end{equation}
This is the key issue of the present Section.

We define the action of the cyclic group of order two, 
 $C_2 \equiv \{\I, P\}$, on the space of the square integrable functions on the line $L^2(\R)$, where $\I$ is the identity and $P$ is the reflection operator,
\begin{equation}\label{2.5}
P f(x) = f(-x)\,,
\end{equation}
with $P^2=\I$. Using  the elements of $C_2$, we can construct the following projectors 
\begin{equation}\label{2.6}
P_E = \frac{1}{2}(\I+P)\,, \qquad P_O = \frac{1}{2}(\I-P) \,,
\end{equation}
that  split  $L^2(\R)$  into two mutually orthogonal subspaces: the spaces of even and odd functions $L^2_E(\R)$ and
$L^2_O(\R)$, respectively, 
\begin{equation}\label{2.7}
 L^2(\R)=L^2_E(\R)\oplus L^2_O(\R)\,.
\end{equation}

Even or odd Hermite functions $\psi_{n}(x)$ are labeled by an even or odd index $n$, respectively. This shows that  \be\label{2.8}
 \psi _n (-x)=(-1)^{n}\,\psi _n (x)\,,\qquad n\in \N\,.
 \ee
 This property, together with the orthogonality of the Hermite functions on $L^2(\mathbb R)$, suggests the following notation
\be\label{2.9}
\{\psi_n(x)\}_{n\in \N} = \{\psi_{2n}(x)\} \oplus \{\psi_{2n+1}(x)\} \,,
\ee
where the analogy with \eqref{2.7} is obvious. 

Next, let us use the property of the orthogonality \eqref{2.2}  of the Hermite functions  as well as  their defined parity \eqref{2.8}, so as to obtain the following identities
\begin{equation}\begin{array}{lll}\label{2.10}
\ds \sum_{n=0}^\infty \psi_n(x)\,\psi_n(y)&=&\ds\sum_{n=0}^\infty \psi_{2n}(x)\,\psi_{2n}(y)+\sum_{n=0}^\infty \psi_{2n+1}(x)\,\psi_{2n+1}(y)=\delta(x-y)\,,\\[0.5cm]
\ds \sum_{n=0}^\infty \psi_n(x)\,\psi_n(-y)&=&\ds\sum_{n=0}^\infty \psi_{2n}(x)\,\psi_{2n}(-y)+\sum_{n=0}^\infty \psi_{2n+1}(x)\,\psi_{2n+1}(-y)
\\[0.4cm]
\ds &=&\ds\sum_{n=0}^\infty \psi_{2n}(x)\,\psi_{2n}(y)-\sum_{n=0}^\infty \psi_{2n+1}(x)\,\psi_{2n+1}(y)=\delta(x+y)\,.
\end{array}\end{equation}
We easily find from the first and the third rows of \eqref{2.10} that
\begin{equation}\begin{array}{lll}\label{2.11}
\ds \sum_{n\in \N} \psi_{2n}(x)\, \psi_{2n}(y) &=&\ds  \frac{1}{2}(\delta(x-y)+\delta(x+y))\,,
\\[0.4cm]
\ds \sum_{n\in \N} \psi_{2n+1}(x)\, \psi_{2n+1}(y) &=&\ds  \frac{1}{2}(\delta(x-y)-\delta(x+y))\,.
\end{array}\end{equation}
Observe that the former and second expressions in \eqref{2.11} gives completeness in the space of the even/odd functions, respectively.

The group $C_2$ is a subgroup of $C_4$, which is  the cyclic group of the four roots of unity $\{1, i, -1, -i\}$.  This comes from two simple facts:  the set of Hermite functions $\{\psi_n(x)\}$ is a basis in $L^2(\R)$
and the Hermite functions are eigenfunctions of the Fourier Transform \eqref{2.3}. From the properties of the imaginary unit we easily obtain that
\be\begin{array}{llllll}\label{2.12}
\text{FT}[\psi_{4n}(x),x,y] &=& +1\, \psi_{4n}(y)\,, \qquad & \text{FT}[\psi_{4n+1}(x),x,y] &=& +i \,\psi_{4n+1}(y)\,, \\[0.4cm] 
\text{FT}[\psi_{4n+2}(x),x,y] &=& -1\, \psi_{4n+2}(y)\,, \qquad &
\text{FT}[\psi_{4n+3}(x),x,y] &=& -i\, \psi_{4n+3}(y)\,.
\end{array}\ee
Consequently, in analogy with \eqref{2.9}, relations \eqref{2.12} yield to this identity having the form of direct sums
\be\label{2.13}
\{\psi_n(x)\}_{n\in \N} = \{\psi_{4n}(x)\} \oplus \{\psi_{4n+1}(x)\} \oplus \{\psi_{4n+2}(x)\} \oplus \{\psi_{4n+3}(x)\}\,.
\ee
Clearly, the subspaces spanned by the Hermite functions $\{\psi_{4n+k}(x)\}$ with $k=0,1,2,3$ are orthonormal to each other. However, we know that each set of Hermite functions given by one of the values $k=0,1,2,3$ cannot fulfil the completeness relation. They satisfy instead
\be\begin{array}{lll}\label{2.14}
\ds \sum_{n\in \N} \psi_{4n}(x) \,\psi_{4n}(y) &=&\ds +\frac {1}{2\sqrt{2\pi}} \,\cos\, (xy) + \frac{1}{4}(\delta(x-y)+\delta(x+y))\,,
\\[0.4cm]\ds 
\sum_{n\in \N} \psi_{4n+1}(x)\, \psi_{4n+1}(y) &=&\ds + \frac {1}{2\sqrt{2\pi}} \,\sin\, (xy)+ \frac{1}{4}(\delta(x-y)-\delta(x+y))\,,
\\[0.4cm]\ds 
\sum_{n\in \N} \psi_{4n+2}(x)\, \psi_{4n+2}(y) &=&\ds -\frac {1}{2\sqrt{2\pi}} \,\cos \, (xy) + \frac{1}{4}(\delta(x-y)+\delta(x+y))\,,
\\[0.4cm]\ds 
\sum_{n\in \N} \psi_{4n+3}(x)\, \psi_{4n+3}(y) &=&\ds -\frac {1}{2\sqrt{2\pi}} \,\sin \, (xy) + \frac{1}{4}(\delta(x-y)-\delta(x+y))\,.
\end{array}\ee

In order to prove relations \eqref{2.14}, let us consider the following Fourier transform (see notation in \eqref{2.3})
\be\label{2.15}\begin{array}{lll}
\ds \text{FT}\left[\sum_{n=0}^N \psi_{n}(x)\, \psi_{n}(y),x,z\right] = \ds\frac 1{\sqrt{2\pi}}  \int_{-\infty}^\infty e^{ixz}\,dx\,\left(\sum_{n=0}^N \psi_{n}(x)\,\psi_{n}(y)\right)\\[0.5cm]\hskip1.8cm 
=\ds  \sum_{n=0}^N \psi_{n}(y)\,\frac {1}{\sqrt{2\pi}}  \int_{-\infty}^\infty e^{ixz}\,\psi_{n}(x)\,dx
= \sum_{n=0}^N i^{n}\,\psi_{n}(y)\,\psi_{n}(z)\,. 
\end{array}\ee
Then, we recall that the second identity in \eqref{2.2} means that, when we endow  the space of the tempered distributions $S'$ with the weak topology where $S$ is the Schwartz space, we have\[
\sum_{n=0}^N \psi_n(x)\psi_n(y)\;\; \xrightarrow{N\to\infty}\;\; \delta(x-y)\,.
\] 
The Fourier transform is weakly continuous on $S'$, hence we have
\begin{eqnarray}\label{2.16}
\ds \text{FT}\left[\sum_{n=0}^N \psi_{n}(x)\, \psi_{n}(y),x,z\right] \;\; \xrightarrow{N\to\infty}\;\; 
 \ds \text{FT}\left[\delta(x-y),x,z\right] = \frac{e^{ixy}}{\sqrt{2\pi}}\,,
\end{eqnarray}
and
\begin{equation}\label{2.17}
\ds \text{FT}\left[\sum_{n=0}^N \psi_{n}(x)\, \psi_{n}(y),x,z\right] \;\; \xrightarrow{N\to\infty}\;\; \sum_{n=0}^{ \infty} i^{n}\,\psi_{n}(x)\,\psi_{n}(y)\,.
\end{equation}
The uniqueness of the weak limit in $S'$ gives
\begin{equation}\label{2.18}
 \sum_{n\in \N} i^{n}\,\psi_{n}(x)\,\psi_{n}(y)
=\frac {e^{ixy}}{\sqrt{2\pi}} \,.
\end{equation}
Obviously, \eqref{2.18} yields to
\be\begin{array}{lll}\label{2.19}
\ds \sum_{n\in \N} \psi_{4n}(x)\,\psi_{4n}(y)-\sum_{n\in \N}  \psi_{4n+2}(x)\,\psi_{4n+2}(y)
\\[0.4cm]\hskip1cm \ds
+\sum_{n\in \N} i\,\psi_{4n+1}(x)\,\psi_{4n+1}(y)-\sum_{n\in \N} i\,\psi_{4n+3}(x)\,\psi_{4n+3}(y)
=\frac {e^{ixy}}{\sqrt{2\pi}} \,.
\end{array}\ee
After the transformation $y\to -y$, we use the parity property of  Hermite functions given in~\eqref{2.8}, so that \eqref{2.19} gives
\be\begin{array}{lll}\label{2.20}
\ds \sum_{n\in \N} \psi_{4n}(x)\,\psi_{4n}(y)-\sum_{n\in \N}  \psi_{4n+2}(x)\,\psi_{4n+2}(y)
\\[0.4cm]\hskip1cm \ds
-\sum_{n\in \N} i\,\psi_{4n+1}(x)\,\psi_{4n+1}(y)+\sum_{n\in \N} i\,\psi_{4n+3}(x)\,\psi_{4n+3}(y)
=\frac {e^{-ixy}}{\sqrt{2\pi}} \,.
\end{array}\ee
Summing   \eqref{2.19} and  \eqref{2.20}, we obtain
\be\begin{array}{lll}\label{2.21}
\ds \sum_{n\in \N} \psi_{4n}(x)\,\psi_{4n}(y)-\sum_{n\in \N}  \psi_{4n+2}(x)\,\psi_{4n+2}(y)
=\frac {e^{ixy}+e^{-ixy}}{2\sqrt{2\pi}}=\frac {1}{\sqrt{2\pi}} \,\cos\, (xy)\,.
\end{array}\ee
Subtracting \eqref{2.20} from   \eqref{2.19}, we obtain that
\be\begin{array}{lll}\label{2.22}
\ds \sum_{n\in \N} \psi_{4n+1}(x)\,\psi_{4n+1}(y)-\sum_{n\in \N}  \psi_{4n+3}(x)\,\psi_{4n+3}(y)
=\frac {e^{ixy}-e^{-ixy}}{2i\sqrt{2\pi}}=\frac {1}{\sqrt{2\pi}} \,\sin\, (xy)\,.
\end{array}\ee
Let us go back to \eqref{2.8}, which can be now rewritten as
\begin{equation}\begin{array}{lll}\label{2.23}
\ds  \sum_{n\in \N} \psi_{4n}(x)\,\psi_{4n}(y)+\sum_{n\in \N}  \psi_{4n+2}(x)\,\psi_{4n+2}(y) 
&=&\ds  \frac{1}{2}(\delta(x-y)+\delta(x+y))\,,
\\[0.4cm]
\ds \sum_{n\in \N} \psi_{4n+1}(x)\,\psi_{4n+1}(y)+\sum_{n\in \N}  \psi_{4n+3}(x)\,\psi_{4n+3}(y)
 &=&\ds  \frac{1}{2}(\delta(x-y)-\delta(x+y))\,.
\end{array}\end{equation}
Finally, let us sum \eqref{2.21} with the first identity of \eqref{2.23}. The result is 
\be\begin{array}{lll}\label{2.24}
\ds \sum_{n\in \N} \psi_{4n}(x)\,\psi_{4n}(y)
=\frac {1}{2\sqrt{2\pi}} \,\cos\, (xy) + \frac{1}{4}(\delta(x-y)+\delta(x+y))\,.
\end{array}\ee
This is the first identity in \eqref{2.14}. The validity of the other three relations in \eqref{2.14} is proven analogously. The proof is now complete.

Next, let us consider an arbitrary function $f(x)$ in $L^2(\R)$. It allows  an expansion in terms of the Hermite functions $\psi_{n}(x)$ as 
\be\label{2.25}
f(x) = \sum_n f_{n} \,\psi_{n}(x)\,,\qquad
f_{n} = \int_{-\infty}^{+\infty} dx\, f(x)\, \psi_{n}(x)\,.
\end{equation}
The even part of $f(x)$, $f_E(x)$,  allows an expansion in terms of even Hermite functions.~Then,
\begin{equation}\label{2.27}
f_{E}(x) = \sum_n f_{2n} \psi_{2n}(x), \qquad f_{2n} = \int_{-\infty}^{+\infty} dx\, f_{}(x) \,\psi_{2n}(x)
\end{equation}
and taking into account  \eqref{2.9}, we write $f_E(x) = f_{+1}(x) + f_{-1}(x)$ so that
\be\begin{array}{llllll}\label{2.28}
\ds f_{+1}(x) & =& \ds\sum_n f_{4n} \,\psi_{4n}(x), \qquad &f_{4n} &=&\ds \int_{-\infty}^{+\infty} dx 
\,f_{}(x)\, \psi_{4n}(x)\,,\\[0.4cm]
\ds f_{-1}(x) &=& \ds \sum_n f_{4n+2} \,\psi_{4n+2}(x), \qquad &f_{4n+2} &=&\ds \int_{-\infty}^{+\infty} dx \,f_{}(x)\, \psi_{4n+2}(x)\,.
\end{array}\ee
For the odd part, $f_O(x)$, we have
\begin{equation}\label{2.29}
f_{O}(x) = \sum_n f_{2n+1} \psi_{2n+1}(x), \qquad f_{2n+1} = \int_{-\infty}^{+\infty} dx\, f_{}(x) \,\psi_{2n+1}(x)\,.
\end{equation}
In analogy with $f_E(x)$, let us split $f_O(x)$  as $f_O(x) = f_{+i}(x) + f_{-i}(x)$, where
\begin{equation}\begin{array}{llllll}\label{2.30}
\ds f_{+i}(x) &=& \ds\sum_n f_{4n+1} \,\psi_{4n+1}(x), \qquad &f_{4n+1} &=&\ds \int_{-\infty}^{+\infty} dx\, f_{}(x)\, \psi_{4n+1}(x)
\,,\\[0.4cm]
\ds f_{-i}(x) &=&\ds \sum_n f_{4n+3} \,\psi_{4n+3}(x), \qquad &f_{4n+3} &=&\ds \int_{-\infty}^{+\infty} dx\, f_{}(x)\, \psi_{4n+3}(x)\,.
\end{array}\end{equation}
All the above results together show that any
 function $f(x)$ in $L^2(\R)$ can  be split into four parts
\be\label{2.31}
f(x) = f_{+1}(x) + f_{+i}(x) + f_{-1}(x) + f_{-i}(x)\,,
\ee
so that if $g(x):= \text{FT}[f(y),y,x]$,
\be\begin{array}{lll}\label{2.32}
\ds f_{+1}(x) = \sum f_{4n}\, \psi_{4n}(x) &= &\ds\frac{1}{4}\left(f(x)+f(-x)+g(x)+g(-x)\right)\,,\\[0.4cm]
\ds f_{+i}(x) = \sum f_{4n+1}\, \psi_{4n+1}(x) &=&\ds \frac{1}{4}\left(f(x)-f(-x)-i g(x)+i g(-x)\right)\,,\\[0.4cm]
\ds f_{-1}(x) = \sum f_{4n+2}\, \psi_{4n+2}(x) &= &\ds\frac{1}{4}\left(f(x)+f(-x)-g(x)-g(-x)\right)\,,\\[0.4cm]
\ds f_{-i}(x) = \sum f_{4n+3}\, \psi_{4n+3}(x) &= &\ds\frac{1}{4}\left(f(x)-f(-x)+i g(x)-i g(-x)\right)\,.
\end{array}\ee
Then, from \eqref{2.25} and the parity of the Hermite functions, it comes that the  even part of $f(x)$ has the following form
\be\begin{array}{lll}\label{2.33}
f_E(x)=\frac{f(x)+f(-x)}{2} &=&\ds \sum_{n\in \N} f_{2n}\,\psi_{2n}(x)\\[0.5cm]
&=&\ds \sum_{n\in \N} f_{4n}\,\psi_{4n}(x)+\sum_{n\in \N} f_{4n+2}\,\psi_{4n+2}(x)\,.
\end{array}\ee
Next, using the continuity of the Fourier transform in $L^2(\mathbb R)$ and taking into account \eqref{2.25}, we have
\begin{equation}\begin{array}{lll}\label{2.34}
g(x)=\text{FT}[f(y),y,x]&=&\ds \frac 1{\sqrt{2\pi}}  \int_{-\infty}^\infty e^{ixy}\,f(y)\,dy\\[0.5cm]
&=&\ds \frac 1{\sqrt{2\pi}}  \int_{-\infty}^\infty e^{ixy}\,dy\,\left(\sum_{n\in \N} f_{n}\,\psi_{n}(y)\right)\\[0.5cm]
&=&\ds  \sum_{n\in \N} f_{n}\,\frac 1{\sqrt{2\pi}}  \int_{-\infty}^\infty e^{ixy}\,\psi_{n}(y)\,dy\\[0.5cm]
&=&\ds \sum_{n\in \N} i^{n}\,f_{n}\,\psi_{n}(x)\,.
\end{array}\end{equation}
Hence,
\be\begin{array}{lll}\label{2.35}
\ds g_E(x)=\frac{g(x)+g(-x)}{2}&=&\ds\sum_{n\in \N}  i^{2n}\,f_{2n}\,\psi_{2n}(x)\\[0.5cm]
&=&\ds\sum_{n\in \N} f_{4n}\,\psi_{4n}(x)-\sum_{n\in \N} f_{4n+2}\,\psi_{4n+2}(x)\,.
\end{array}\ee
From \eqref{2.33} and  \eqref{2.35}, we obtain the first expression of  \eqref{2.32}. Similarly, we prove all the other relations in  \eqref{2.32}.
The projectors producing this splitting are
\begin{equation}\begin{array}{ll}\label{2.36}
P_{+1} = \frac{1}{4}(1+P)(1+\text{FT})\,, \qquad & P_{+i} = \frac{1}{4}(1-P)(1-i\,\text{FT})\,,\\[0.4cm]
 \;P_{-1} = \frac{1}{4}(1+P)(1-\text{FT})\,, \qquad &P_{-i} = \frac{1}{4}(1-P)(1+i\,\text{FT})\,.
\end{array}\end{equation}
They verify the following identities
\begin{equation}\label{2.37}
P_{+1} + P_{+i} + P_{-1} + P_{-i}=\I , \quad P_E = P_{+1}+P_{-1}\,, \quad P_O = P_{+i} + P_{-i}\,.
\end{equation}
All these projections are orthogonal and the corresponding splitting of the Hilbert space $L^2(\mathbb R)$ is given by
\begin{equation}\label{2.38}
 L^2(\R)=L^2_E(\R)\oplus L^2_O(\R)=L^2_{+1}(\R)\oplus L^2_{-1}(\R)\oplus L^2_{+i}(\R)\oplus L^2_{-i}(\R)\,.
\end{equation}

We conclude this part of the discussion here. 

\section{Periodic Functions 
and Hermite Functions}\label{periodicfunctions}

As is well known and as we have mentioned before,  the complete set of Hermite functions, $\{\psi_n(x)\}_{n\in\N}$, forms an orthonormal basis on
$L^2(\mathbb R)$. Next, and  using the Hermite functions, we shall construct a countable set of periodic functions that will be a system of generators of the space of square integrable functions on the unit circle. 

The space $L^2(\mathcal C)$ is the space of Lebesgue square integrable functions $f(\phi):\mathcal C\longmapsto \mathbb C$. The norm  of the function $f(\phi)$ is given by the following relation
\begin{equation}\label{3.1}
 ||f(\phi)||^2:=\frac{1}{2\pi}\,\int_{-\pi}^{\pi}|f(\phi)|^2\,d\phi<\infty\,.
\end{equation}

We intend to introduce a space of functions with given properties.  Consider the angular variable $-\pi\le \phi<\pi$ and  define
\begin{equation}\label{3.2}
\c_n(\phi):= \sum_{k=-\infty}^{+\infty} \psi_n(\phi+2k\pi)\,,\qquad k\in\Z\,,\; n=0,1,2,\dots\,.
\end{equation}
The functions $\c_n(\phi)$ are obviously periodic with period $2\pi$
\begin{equation}\label{3.3}
\c_n(\phi+2\pi)=\sum_{k=-\infty}^{\infty} \psi_n(\phi+2(k+1)\pi)=\sum_{k=-\infty}^\infty \psi_n(\phi+2k\pi)=\c_n(\phi)\,.
\end{equation}

A first result about the convergence of the series defining $\c_n(\phi)$ is given by the 
 following proposition.\medskip

{\bf Proposition 1.-} {\sl 
The series defining each of the $\c_n(\phi)$ are absolutely convergent. Furthermore, each $\c_n(\phi)$ is bound 
 on the interval $-\pi\le\phi<\pi$ and the square integrable on this interval. 
}\medskip

{\bf Proof}
Let us write \eqref{3.2} as
\begin{equation}\label{3.4}
\c_n(\phi) =\psi_0(\phi)+\sum_{k=1}^\infty \psi_n(\phi+2k\pi)+\sum_{k=-\infty}^{-1} \psi_n(\phi+2k\pi)\,.
\end{equation} 
The first term in the right hand side of \eqref{3.3} is bound by  $\psi_0(\phi)\leq 2\,, \forall \phi \in [-\pi,\pi) $, as we may check by using the Cram\'er inequality \cite{abramowitz} (page 787, formula 22.14.17). Since both series in \eqref{3.4} are similar, it is sufficient to analyze one of them, as the conclusions for the other one would be the same. Let us consider the series with $k\ge 1$.  Then, we have to study the convergence of the series
\begin{equation}\label{3.5}
\left| \sum_{k=1}^\infty \psi_n(\phi+2k\pi)  \right| \le \sum_{k=1}^\infty \big|\psi_n(\phi+2k\pi)\big|\,.
\end{equation}
From \eqref{2.1} we have that
\begin{equation}\label{3.6}\ds
\psi_n(\phi+2k\pi)=  \frac 1{\pi^{1/4}\,2^{n/2}\,\sqrt{n!}}\,e^{-(\phi+2k\pi)^2/2}\,H_n(\phi+2k\pi)\,,
\end{equation}
with $\phi+2k\pi\geq 1 \,, \forall k=1,2,\dots$.
For any real value of $x$, $H_n(x)$ is a real polynomial of order $n$ with the property that for $|x|>1$ has the following upper bound 
\begin{equation}\label{3.7}
\big| H_n(x) \big|\le 2^n\,(n+1)!\,\big| x\big|^n\,,
\end{equation}
{which  is a straightforward consequence of} the know formula of the Hermite  polynomials, 
{which states that}
\begin{equation}\label{3.8}
H_n(x)= n!\,\sum_{m=0}^{\lfloor n/2\rfloor}\frac{(-1)^m}{m!\,(n-2m)!}\, (2x)^{n-2m}\,.
\end{equation}
Consequently, since $\pi^{1/4}\,2^{n/2}\,\sqrt{n!}\geq 1,\, \forall n\in \N$, we have from \eqref{3.6} and \eqref{3.7} an upper bound for  $\big| \psi_n(\phi+2k\pi) \big|$
\be \begin{array}{lll}\label{3.9}
\ds
\big|\psi_n(\phi+2k\pi)\big| &\le& 2^n\,(n+1)!\, e^{-(\phi+2k\pi)^2/2}\,(\phi+2k\pi)^n 
 \\[2ex]
&\le &  2^n\,(n+1)!\, e^{-(\phi+2k\pi)/2}\,(\phi+2k\pi)^n \\[2ex]
&\le &  2^n\,(n+1)!\, e^{-k\pi}\, [2(k+1)\pi]^n  \\[2ex]
&\le &   2^n\,(n+1)!\,(2\pi)^n \,e^{\pi} \, e^{-(k+1)\pi}\,(k+1)^n \\[2ex]
&\le &  (2\pi)^{2 n} \,e^{\pi} \, (n+1)!\,e^{-(k+1)}\, (k+1)^n\,.
\end{array}\ee
Hence, 
since $n$ is fixed and after \eqref{3.9}, the sum \eqref{3.5} converges if the following series
\be\label{3.10}
\sum_{k=1}^\infty  e^{-(k+1)}\, (k+1)^n
\ee 
 converges. This convergence  
  is a simple exercise of analysis.

The conclusion is that the series in \eqref{3.4} with $k\ge 1$ is absolutely convergent and hence pointwise convergent. In particular, this means that the functions $\mathfrak C_n$ are all Lebesgue measurable, since they are the pointwise limit of measurable functions.   Similarly, the same property is valid for the series with $k<-1$  in \eqref{3.4}. This shows the boundedness of $\c_n(\phi)$ on $-\pi\le\phi<\pi$ for $n=0,1,2,\dots$, which, along with its measurability in the Lebesgue sense, shows that the functions $\c_n(\phi)$ are square integrable in the considered interval.
\hfill $\square$\medskip

Proposition 1 has an important consequence. We have shown that, for each $n\in\N$, the function $\c_n(\phi)$ has an upper bound. 
 The constant function that equals to this upper bound 
in the interval $[-\pi,\pi)$ is square integrable, so that by using the Lebesgue dominated convergence theorem \cite{AB} we have that
\begin{eqnarray}\label{3.11}
\int_{-\pi}^{\pi} e^{im\phi}\,d\phi \sum_{k=-\infty}^\infty \psi_n(\phi+2k\pi) =\sum_{k=-\infty}^\infty \int_{-\pi}^{\pi} e^{im\phi}\, \psi_n(\phi+2k\pi)\,d\phi\,.
\end{eqnarray} 
We also have that
\be\label{3.12}
\text{FT}[\c_n(x),x,y ]=\sum_{k=-\infty}^\infty \text{FT}[\psi_n(x+2k\pi),x,y]\,.
\ee

Finally, we may  mention another interesting fact: the functions $\{\c_n(\phi)\}_{n\in\N}$ span $L^2[-\pi,\pi)$. This means that the subspace of all finite linear combinations of these functions is dense in $L^2[-\pi,\pi)$. The proof will be given later. \medskip

One of the objectives of the present article is to find some relations between functions that are of use in Fourier analysis. We shall discuss this idea along the next Section.


\section{Fourier Transform on the Circle}\label{FTcircle}

After the general formula for the Fourier Transform given in  \eqref{2.4}, we obtain that
\begin{equation}\begin{array}{lll}\label{4.1}
\text{FT}[\psi_n(x+a)+\psi_n(x-a),x,y] &=&\text{FT}[\psi_n(x+a),x,y]+ \text{FT}[\psi_n(x-a),x,y] \\[0.4cm]
&=& 2 i^n \cos \, (ay)\, \psi_n(y)\,,\qquad a\in \R.
\end{array}\end{equation}
Furthermore, the Inverse Fourier Transform (IFT) gives the following relation
\begin{equation}\label{4.2}
\text{IFT}[2 \cos \, (ay)\, \psi_n(y),x] = (-i)^n (\psi_n(x+a)+\psi_n(x-a))\,.
\end{equation}
For the special case  $a=2\pi$, Equations \eqref{4.1} and \eqref{4.2} give, respectively, 
\begin{equation}\begin{array}{rll}\label{4.3}
\text{IFT}[\psi_n(x+2\pi)+\psi_n(x-2\pi),x,y] &=& 2 i^n \cos \,(2\pi y) \,\psi_n(y) 
\,,\\[0.4cm]
\text{IFT}[2 \cos2\pi y\, \psi_n(y),y,x] &=& (-i)^n(\psi_n(x+2\pi)+\psi_n(x-2\pi)) \,.
\end{array}\end{equation}

We may extend the previous formulae for a finite sum of Hermite functions, so that 
\be\begin{array}{lll}\label{4.4}
\ds \text{FT}\left[\sum_{k=-m}^{+m}\psi_n(x+ka),x,y\right] &=& \ds i^n \left(1+2\sum_{k=1}^{m}\cos(kay)\right) \psi_n(y) 
\\[0.5cm] 
&=& i^n \,D_m(ay) \,\psi_n(y)
\,,\\[0.5cm]
\ds \text{FT}\left[\sum_{k=-m}^{+m}\psi_n(x+2\pi k),x,y\right]  &=& i^n\, D_m(2\pi y)\, \psi_n(y)     \,,  
\end{array}\ee
where $D_m(ay)$ is the Dirichlet kernel \cite{dirichlet}, i.e., 
\begin{equation}\label{4.5}
D_m(a y) =\sum_{k=-m}^{m}e^{ i\, k\, a \,y} = \frac{\sin[(m+1/2)\,a\,y ] }{\sin[(1/2)\,a\,y]}.
\end{equation}

At this step, let us introduce the following sequence of functions depending on the natural parameter $m$
\be\label{4.6}
\c_n(x;m):= \sum_{k=-m}^{+m} \psi_n(x+2\pi k)
\ee 
so that $\ds \c_n(x)= \lim_{m\to\infty} \c_n(x;m)$.  
The second equation in  \eqref{4.4} may be rewritten as  
\be\label{4.7}
\text{FT}[\c_n(x;m),y] = i^n\, D_m(2\pi y)\, \psi (n,y)\,.
\ee
From \eqref{3.11}, we conclude that
\begin{equation}\label{4.8}
\text{FT}[\c_n(x),y] = i^n\, \left(\lim_{m\to\infty} D_m(2\pi y)\right) \psi_n(y)\,,
\end{equation}
where
\begin{equation}\label{4.9}
\lim_{m\to\infty} D_m(a y) = \frac{2\pi}{a}\sum_{m=-\infty}^{+\infty} \delta \left(y-\frac{2\pi}{a}m\right)
=\Sha\left(\frac{y}{2\pi/a}\right)\,.
\end{equation}
In particular, for $a=2\pi$ \eqref{4.9} looks like
\begin{equation}\label{4.10}
\lim_{m\to\infty} D_m(2\pi y) = \sum_{m=-\infty}^{+\infty} \delta \left(y-m\right)\equiv  \Sha(y)\,.
\end{equation}
The infinite sum $ \Sha$ is called the Dirac comb \cite{schwartz,cordoba,brandwood}
\[
\Sha_L(y)=  \sum_{m=-\infty}^{+\infty} \delta \left(y-m L\right)= 
\frac{1}{L}\,\sum_{m=-\infty}^{+\infty} \delta \left(y/L-m\right)=\frac{1}{L}\,\Sha \left(y/ L\right)\,.
\]
It converges in the weak sense as a distribution on the space, $\mathcal D$, of all Schwartz functions with compact support on $\mathbb R$ endowed with 
the strict inductive limit topology. For the properties of the Fourier transform we can write that
\be\label{4.11}
\text{FT}[\c_n(x),y] = i^n\, \Sha(y)\,\psi_n(y)\,.       
\ee
As is well known, the Dirac Comb is autoconjugate under Fourier transform.
This is evident~because
\begin{equation}\begin{array}{lll}\label{4.12}
\ds \text{FT}\left[\frac{\sin[(k+1/2)y]}{\sin[(1/2)y]},x\right] &=&\ds  \sqrt{2\pi} \sum_{m=-k}^{+k} \delta(x-m)\,,\\[0.5cm]  
\ds \text{FT}\left[\sum_{m=-k}^{+k} \delta(x-m),y\right] &=& \ds \frac{1}{\sqrt{2\pi}} \frac{\sin[(k+1/2)y}{\sin [(1/2)y]}\,,
\end{array}\end{equation}
and both the Dirichlet kernel and the r.h.s. of the first expression go to the Dirac Comb for $k\to \infty$.

Moreover, the functions  $\{\c_{n}(\phi)\}_{n\in \N}$ can be split first into even and odd and then into four subspaces. Indeed
\be\begin{array}{lll}\label{4.13}
\{\c_{n}(\phi)\}_{n\in \N}&=&\{\c_{2n}(\phi)\}\oplus  \{\c_{2n+1}(\phi)\}\,,\\[0.4cm]
&=&\{\c_{4n}(x)\}\oplus 
\{\c_{4n+1}(\phi)\}\oplus\{\c_{4n+2}(\phi)\}\oplus \{\c_{4n+3}(\phi)\}\,.
\end{array}\ee

\section{A Discretized Fourier Transform}\label{discretefourier}

To begin with,  let us compare the space $L^2(\mathcal C)$, also denoted as $L^2[-\pi,\pi)$, with the space $l_2(\Z)$.   As is well known, an orthonormal basis on $L^2(\mathcal C)$ is $\{(2\pi)^{-1}\,e^{im\phi}\}_{m\in \Z}$, hence
\begin{equation}\label{5.1}
f(\phi)=\frac1{2\pi} \sum_{m\in\mathbb Z}f_n\,e^{-im\phi}\,,\qquad f\in L^2(\mathcal C)\,,
\end{equation} where the sum converges in the sense of the norm \eqref{3.1} (for continuous functions $f(\phi)$ the series also converge pointwise \cite{TYN}). The properties of orthonormal bases in Hilbert spaces show that 
\begin{equation}\label{5.2}
\sum_{m\in\mathbb Z}\big| f_m\big|^2=2\pi\, \big|\big| f(\phi) \big|\big|^2\,.
\end{equation}
We  call to the complex numbers $f_m$  ($m\in\mathbb Z$) the components of $f$.

The Hilbert space $l_2(\mathbb Z)$ is a space of sequences of complex numbers $A\equiv\{a_m\}_{m\in\mathbb Z}$ such that
\begin{equation}\label{5.3}
\big|\big| A \big|\big|^2:=\sum_{m\in\mathbb Z}\big| a_m\big|^2<\infty\,.
\end{equation}
This is a Hilbert space with a scalar product given by
\begin{equation}\label{5.4}
( A,B):= 
\frac1{2\pi}\,
 \sum_{m\in\mathbb Z} a^*_m\,b_m\,,
\end{equation}
where  the star denotes complex conjugation.

An orthonormal basis for $l_2(\Z)$ is given by the sequences that have all their components equal to zero except for one which is equal to one. Let us call $\{\mathcal B_m\}_{m\in\mathbb Z}$ on this basis,  
where each of the $\mathcal B_m$ represents each one of these series. Any $F\in l_2(\Z)$ with components  
$F\equiv \{f_m\}_{m\in\Z}$ may be written as
\begin{equation}\label{5.5}
F=\frac1{2\pi}\,
\sum_{m\in\Z} f_m\,\mathcal B_m\,, \qquad {\rm with} \qquad
 \sum_{m\in\Z}\big| f_m\big|^2=2\pi\,
  \big|\big|F \big|\big|^2<\infty\,.
\end{equation}

We readily see that there exists a correspondence between $L^2(\mathcal C)$ and $l_2(\Z)$. This correspondence relates any $f(\phi)\in L^2(\mathcal C)$ as in \eqref{5.1} with $F'=F/2\pi$ as in \eqref{5.5} with the same sequence $\{f_m\}_{m\in\Z}$. This correspondence, $f(\phi)\longmapsto F$, is clearly linear, one to one and so on. 
 In addition, $ \big|\big|f(\phi) \big|\big|= \big|\big|F \big|\big|$, which shows that it is, in addition, unitary (in fact any pair of infinite dimensional separable Hilbert spaces are unitarily equivalent in the sense that one may construct one, in fact infinite, unitary mappings from one to the other).  
Equation \eqref{5.1} gives the Fourier series span for $f(\phi)\in L^2(\mathcal C)$. From this point of view, we may say that the Fourier series is a unitary mapping, $\mathcal F$, from $L^2(\mathcal C)$ onto $l_2(\Z)$. It admits an inverse, $\mathcal F^{-1}$,  from $l_2(\Z)$ onto $L^2(\mathcal C)$, which is also unitary and is sometimes called the {discrete Fourier transform} \cite{oppenheim}. 

We intend to offer a homogeneous version of concepts that are often introduced as separated. They are the {Fourier transform}, {Fourier series}, and {discrete Fourier transform} on one side and the {Hermite functions} on the other. 

The first step is to construct a set of sequences in $l_2(\Z)$ using the Hermite functions $\psi_n(x)$.   
For each $n\in \N$, let us define the following sequence indexed by the set of integer numbers $\mathbb Z$
\begin{equation}\label{5.6}
\chi_n:= \{\psi_n(m)\}_{m\in\mathbb Z}\,.
\end{equation}

Our next result has not been proven so far. Its proof requires the previous demonstration of a couple of results displayed in Appendix  \ref{apendb} (Propositions A1 and A2).

Since the functions $\c_n(\phi)$ are in $L^2[-\pi,\pi)$, they admit a span in terms of the orthonormal basis in $L^2[-\pi,\pi)$ as commented upon at the beginning of the present Section (see Equation~\eqref{5.1}). Thus, we can write
\begin{equation}\label{5.7}
\c_n(\phi)= \frac{1}{\sqrt{2\pi}} \sum_{m=-\infty}^\infty c_n^m\,e^{-im\phi}\,,
\end{equation}
with
\begin{equation}\label{5.8}
c_n^m= \frac{1}{\sqrt{2\pi}} \int_{-\pi}^\pi e^{im\phi}\,\c_n(\phi)\,d\phi\,.
\end{equation}
The continuity of the functions  $\c_n(\phi)$ on $[-\pi,\pi)$ guarantees the pointwise convergence of \eqref{5.7} \cite{TYN}. In addition, since all $\c_n(\phi)$ are periodic with period $2\pi$, \eqref{5.7} is valid for all real numbers of $\phi$. 
We recall that each of the Hermite functions $\psi_n(x)$ are eigenfunctions of the Fourier Transform with eigenvalue $i^n$.  Let us use this idea in \eqref{5.8} in order to find an explicit expression of the coefficients $c_n^m$ in terms of the values of the Hermite functions at the integers. Using the definition \eqref{3.2} of the $\c_n(\phi)$ in \eqref{5.8}, we obtain
\be\begin{array}{lll}\label{5.9}
c_n^m &=& \ds
\frac{1}{\sqrt{2\pi}} \int_{-\pi}^\pi e^{im\phi}\,d\phi \left[ \sum_{k=-\infty}^\infty \psi_n(\phi+2k\pi) \right]
\\[0.4cm] &=&\ds \frac{1}{\sqrt{2\pi}} \sum_{k=-\infty}^\infty \int_{-\pi}^\pi e^{im\phi}\, \psi_n(\phi+2k\pi)\,d\phi  \\[0.4cm] 
&=& \ds 
\frac{1}{\sqrt{2\pi}} \sum_{k=-\infty}^\infty \int_{-\pi+2k\pi}^{\pi+2k\pi}  e^{ims}\,\psi_n(s)\,ds \\[0.4cm]
&=&\ds \frac{1}{\sqrt{2\pi}} \int_{-\infty}^\infty  e^{ims}\,\psi_n(s)\,ds = i^n\,\psi_n(m)\,.
\end{array}\ee
The second identity in \eqref{5.9} makes use of the Lebesgue-dominated convergence theorem~\cite{AB} in order to interchange the integral and the series as anticipated in \eqref{2.15}. We have also used the change of variable $s=\phi+2k\pi$ and $e^{im\phi}=e^{im(\phi+2k\pi)}=e^{ims}$. This shows that \eqref{5.7} and \eqref{5.8} can be written, respectively, as
\begin{equation}\label{5.10}
\c_n(\phi)= \frac{i^n}{\sqrt{2\pi}} \sum_{m=-\infty}^\infty \psi_n(m)\,e^{-im\phi}
\end{equation}
and
\begin{equation}\label{5.11}
\psi_n(m)= \frac{(-i)^n}{\sqrt{2\pi}} \int_{-\pi}^\pi  \c_n(\phi)\,e^{im\phi}\,d\phi\,.
\end{equation}
After  \eqref{5.10} and  \eqref{5.11}, we may obtain a relation between the functions $\c_n(\phi)$ and the  series  $\chi_n$ defined in \eqref{5.6} as
\be \label{5.12}
\c_n(\phi)\equiv i^n\, \chi_n\,.
\ee
Thus, \eqref{5.10} and \eqref{5.11} provide a one to one relation between the functions $\c_n(\phi)$, as defined in \eqref{3.2} and the sequences $\chi_n$ given in \eqref{5.6}.

Observe that the definition  \eqref{3.2} of $\c_n(\phi)$ allows us to write \eqref{5.10} as
\begin{equation}\label{38}
\sum_{m=-\infty}^\infty \left[ \psi_n(\phi+2m\pi) - \frac{i^n}{\sqrt{2\pi}}\,\psi_n(m)\,e^{-im\phi}  \right]=0\,,\qquad n=0,1,2,\dots\,.
\end{equation}

So far, we have discussed the relation between a system of generators in $L^2[-\pi,\pi)\equiv L^2(\mathcal C)$ given by $\{\c_n(\phi)\}_{n\in \N}$ and a set of series $\{\chi_n\}_{n\in \N}$ in $l_2(\mathbb Z)$. These systems of generators do not form  orthonormal bases on   $L^2(\mathcal C)$ and $l_2(\mathbb Z)$, respectively. 


\section{Relevant Operators Acting on $L^2(\mathcal C)$ and on $l^2(\Z)$}\label{relevantoperators}

We want to discuss some of the properties of the functions $\{\c_n(\phi)\}$ in relation with its behavior under different operators.
Here, we introduce a set of operators on $L^2(\mathcal C)$ and on $l^2(\Z)$ similar to  the operators acting on the quantum harmonic oscillator. In particular, we have creation and annihilation and number operators that act on the chosen basis as expected. In addition, we have some other operators which play the role of multiplication by the variable and differentiation, with the expected relation with the ladder operators. In the present context, the definitions of these operators has to be done on a particular form. As also happens in relation to the harmonic oscillator, these operators are not bound on the Hilbert spaces they act on and have different domains.  Nevertheless, we may equip these Hilbert spaces with a dual pair of locally convex spaces, so that these operators be continuous. This construction will be done in the next Section.  

The possibility of introducing multiplication and differentiation operators on the circle has been previously considered and noticed that serious inconsistencies emerge when we try to extend these operators to the circle on its most natural form \cite{CN,ZA,ZA1,LE,MT,kastrup,gazeau16,gazeau18}. In particular, the need for boundary conditions for the wave functions of the form $\varphi(0)=e^{iak}\,\varphi(0)$ produce an ambiguity on the definition of the derivation operator, which now has to depend on the parameter $k$. In the formalism we introduced in the sequel, which in part has been based on the Weil--Brezin--Zak transformation \cite{ZA,JA,FO}, we try to avoid some of these inconveniences \cite{CN,ZA,LE}.

\subsection{Multiplication and Derivation Operators on $L^2(\mathcal C)$}
\label{sec6.1}

 Let us give some definitions such as multiplication and derivation operators within our context. 
 
 \subsubsection{Multiplication Operator}
  
 For the multiplication operator $\Phi$, we have to discard the apparently most natural definition that, for any $f(\phi)$, it would have been  $\Phi f(\phi)=\phi\,f(\phi)$.  Since $f(\phi)$ is a function on the circle that should be extended by periodicity to the real line, this definition is not appropriate as it does not provide a periodic function. Let us define the operator $\Phi$ by means of its action on each of the functions of the sequence $\{\c_n(\phi)\}$ and, then, extend it by linearity. Therefore, $\Phi$ would be defined on a dense set of the closed subspace spanned by the vectors of the sequence $\{\c_n(\phi)\}$. Our definition is
\begin{equation}\label{6.1}
\Phi\,\c_n(\phi):= \sum_{k=-\infty}^\infty (\phi+2k\pi)\,\psi_n(\phi+2k\pi)\,,
\end{equation}
which is indeed periodic with period $2\pi$. We may extend this definition for any real $a\ge 1$~as
\begin{equation}\label{6.2}
\Phi^a\,\c_n(\phi):= \sum_{k=-\infty}^\infty (\phi+2k\pi)^a\,\psi_n(\phi+2k\pi)\,,
\end{equation}
since the series in the r.h.s. in \eqref{6.2} are absolutely convergent.

We want to study some properties of the operator $\Phi$. We begin with the following property valid for Hermite polynomials
\begin{equation}\label{6.3}
xH_n(x)=\frac 12\,H_{n+1}(x)+n\,H_{n-1}(x)\,.
\end{equation}
Using this property in \eqref{6.1}, we obtain
\be\begin{array}{lll}\label{6.4}
\Phi \,\c_n(\phi)&=&\ds \sum_{k=-\infty}^\infty (\phi+2k\pi)\, H_n(\phi+2k\pi)\,e^{-(\phi+2k\pi)^2/2}  \\[0.4cm]
&=& \ds\frac12\, \sum_{k=-\infty}^\infty H_{n+1}(\phi+2k\pi)\,e^{-(\phi+2k\pi)^2/2} \\[0.4cm]
&& \qquad\ds + n \sum_{k=-\infty}^\infty H_{n-1}(\phi+2k\pi)\,e^{-(\phi+2k\pi)^2/2} \\[0.4cm] 
&=&\ds\frac 12 \,\c_{n+1}(\phi)+n\,\c_{n-1}(\phi)\,.
\end{array}\ee
Note that $\Phi \c_0(\phi)\propto \c_1(\phi)$. 

 \subsubsection{Derivative Operator}
 
 Next, let us define the derivative operator $D_\phi$ on the subspace of all linear combinations of the functions $\{\c_n(\phi)\}_{n\in\N}$. Clearly, we just need to define the action $D_\phi$ on each of these functions. To begin with, let us write
\begin{equation}\label{6.5}
D_\phi\,\c_n(\phi):=\sum_{k=-\infty}^\infty \frac{d\,\psi_n(\phi+2k\pi)}{d\,(\phi+2k\pi)}\,, \qquad n=0,1,2,\dots \,.
\end{equation}
From the properties of the Hermite functions,  
we may obtain two different expressions, although equivalent,  for 
the derivative of a Hermite function,  
which are
\be\begin{array}{lll}
\psi_n'(x)=\sqrt{2n}\,\psi_{n-1}(x)-x\,\psi_n(x)\,,
\label{6.6}\\[2ex]
\psi'_n(x)= x\,\psi_n(x)+\sqrt{2(n+1)}\,\psi_{n+1}(x)\,.
\end{array}\ee
Let us use the first expression of  \eqref{6.5} in \eqref{6.6}. The result is
\be\begin{array}{lll}\label{6.7}
D_\phi\,\c_n(\phi)&=&\ds 
-\sum_{k=-\infty}^\infty \{ (\phi+2k\pi)\psi_n(\phi+2k\pi)-\sqrt{2n}\,\psi_{n-1}(\phi+2k\pi) \}
\\[0.4cm] 
&=&\ds -\Phi\,\c_n(\phi)+\sqrt{2n}\,\c_{n-1}(\phi)\,,
\end{array}\ee
where the last identity on \eqref{6.7} comes from the absolute convergence of the series involved and the definition of $\Phi$. 
 Next, using the second expression of \eqref{6.6}, we have
\be\begin{array}{lll}\label{6.8}
D_\phi\,\c_n(\phi)&=&\ds \sum_{k=-\infty}^\infty \{(\phi+2k\pi)\psi_n(\phi+2k\pi) + \sqrt{2(n+1)}\,\psi_{n+1}(\phi+2k\pi)  \} 
\\[0.4cm] 
&=& \Phi\,\c_n(\phi)+\sqrt{2(n+1)}\,\c_{n+1}(\phi)\,.
\end{array}\ee
We may use either one of the equivalent relations \eqref{6.7} or \eqref{6.8} as the definition of the operator $D_\phi$. 

\subsection{Ladder Operators}
\label{sec6.2}

Let us define ladder operators on the subspace of linear combinations of the elements of $\{\c_n(\phi)\}$ as follows
\begin{equation}\label{6.9}
A^+_\phi\,\c_n(\phi):= \sqrt{n+1}\,\c_{n+1}(\phi)\,,\qquad A^-_\phi\,\c_n(\phi):= \sqrt{n}\,\c_{n-1}(\phi)\,.
\end{equation}
From \eqref{6.7}--\eqref{6.9}, we obviously have that
\begin{equation}\label{6.10}
A^+_\phi=\frac 1{\sqrt{2}}\,(D_\phi-\Phi)\,, \qquad A^-_\phi=\frac1{\sqrt 2}\,(D_\phi+\Phi)\,,
\end{equation}
so that
\begin{equation}\label{6.11}
\Phi=\frac 1{\sqrt 2}\, (A^-_\phi-A^+_\phi)\,,\qquad  D_\phi= \frac 1{\sqrt 2}\, (A^-_\phi+A^+_\phi)\,.
\end{equation}
Exactly as with the quantum harmonic oscillator,  we may define the following number operator
\begin{equation}\label{6.12}
N_\phi:= A^+_\phi\,A^-_\phi\,, \qquad \text{so that}\qquad N_\phi\,\c_n(\phi)=n\,\c_n(\phi)\,,\qquad n=0,1,2,\dots \,. 
\end{equation}
It is also obvious that
\begin{equation}\label{6.13}
\frac 12\, (\Phi^2-D^2_\phi) =N_\phi+\frac 12\,.
\end{equation}
This completes the analogy with the quantum harmonic oscillator. 
All these operators admit closed extensions and are unbound.


\subsection{Operators on $l^2(\Z)$}
\label{sec6.3}

In the previous two subsections we have been  concerned with  functions in and operators on subspaces of 
$L^2(\mathcal C)$. Now, let us find the equivalent objects in $l_2(\mathbb Z)$. 

Using the canonical orthonormal basis $\{\mathcal B_m\}_{m\in\mathbb Z}$ of $l_2(\Z)$  defined in Section \ref{discretefourier}, we  can write after \eqref{5.6}
\begin{equation}\label{6.14}
\chi_n=\sum_{m=-\infty}^\infty \psi_n(m)\,\mathcal B_m\,,\qquad \forall n\in \N\,.
\end{equation}
 We define the following operators on the set of the vectors $\{\chi_n\}_{n\in \N}$ and then extend them by linearity  to the subspace of their linear combinations
\begin{equation}\begin{array}{lll}\label{6.15}
M\,\chi_n &=&\ds M\sum_{m=-\infty}^\infty \psi_n(m)\,\mathcal B_m := \sum_{m=-\infty}^\infty m\,\psi_n(m)\,\mathcal B_m\,,\\[0.4cm] M^a\,\chi_n &:=&\ds \sum_{m =-\infty}^\infty m^a\,\psi_n(m)\,\mathcal B_m\,,
\end{array}\end{equation}
$a$ being any real number greater than or equal to  $1$.  Expressions \eqref{6.15} are well defined and belong to $l_2(\mathbb Z)$.
In fact, after \eqref{5.5} and some analysis, we conclude that
\begin{equation}\label{6.16}
\sum_{m=-\infty}^\infty |m^\alpha\,\psi_n(m)|^2 \le 2^{2 n}\,((n+1)!)^2\, \sum_{m=-\infty}^\infty e^{-m^2}\,m^{2(n+a)}<\infty\,.
\end{equation}

In addition, we may also define a formal derivative on the space of finite linear combinations of the $\{\chi_n\}$ as follows
\begin{equation}\label{6.17}
D\,\chi_n : = \sum_{m=-\infty}^\infty \psi'(m)\,\mathcal B_m\,,\qquad  \psi'(m) :=\frac{d}{dx}\,\psi(x) \bigg|_{x=m}\,.
\end{equation}
Choosing the first expression in \eqref{6.6} and taking into account \eqref{6.16}, we have
\be\begin{array}{lll}\label{6.18}
D\,\chi_n 
 &=& \ds\sum_{m=-\infty}^\infty \{ \sqrt{2n}\,\psi_{n-1}(m)-m\,\psi_n(m) \} \mathcal B_m  \\[0.4cm] 
 &=& \ds \sqrt{2n}  \sum_{m=-\infty}^\infty \psi_{n-1}(m)\,\mathcal B_m - \sum_{m=-\infty}^\infty m\,\psi_n(m)\,\mathcal B_m   \\[0.4cm] 
 &=& \sqrt{2n}\, \chi_{n-1}-M\,\chi_n\,,
\end{array}\ee
where the first identity in \eqref{6.18} makes sense because $\{\mathcal B_m\}$ is an orthonormal 
basis.  
Analogously, using the second row in \eqref{6.7}, we have
\be\begin{array}{lll}\label{6.19}
D\,\chi_n&=&\ds
 \sum_{m=-\infty}^\infty \{m\,\psi_n(m)+\sqrt{2(n+1)}\,\psi_{n+1}(m)\}\mathcal B_m \\[0.4cm] 
 &=& \ds \sum_{m=-\infty}^\infty m\,\psi_n(m) \,\mathcal B_m + \sqrt{2(n+1)} \sum_{m=-\infty}^\infty \psi_{n+1}(m)\, \mathcal B_m \\[0.4cm] 
 &=& \ds M\,\chi_n+\sqrt{2(n+1)}\,\chi_{n+1}\,.
\end{array}\ee
Both Equations \eqref{6.18} and \eqref{6.19} show independently that $D$ is well defined in the space of finite linear combinations of vectors $\{\chi_n\}$ and, therefore, \eqref{6.17} makes sense.  In addition, if we define on the same space the creation, $B^+$, and annihilation, $B^-$,  operators  as
\begin{equation}\label{6.20}
B^+\chi_n :=\sqrt{n+1}\,\chi_{n+1}\,,\qquad B^-\,\chi_n :=\sqrt{n}\,\chi_{n-1}\,,
\end{equation}
we have the following relations
\begin{equation}\label{6.21}
B^+=\frac{\sqrt 2}{2}\,(M-D)\,,\qquad B^-=\frac{\sqrt 2}{2}\,(M+D)\,,
\end{equation}
and
\begin{equation}\label{6.22}
M=\frac{\sqrt 2}{2}\, (B^++B^-)\,,\qquad D=\frac{\sqrt 2}{2}\,(B^--B^+)\,.
\end{equation}
Formulas \eqref{6.20}--\eqref{6.22} are equivalent to \eqref{6.9}--\eqref{6.11}, respectively. Then, we may define the corresponding number operator  as
\begin{equation}\label{6.23}
 N_{\mathbb Z}:= B^+\,B^-\,,
\end{equation}
which on the considered systems of generators gives
\begin{equation}\label{6.24}
 N_{\mathbb Z}\,\chi_n=n\,\chi_n\,.
\end{equation}
These expressions give a harmonic oscillator like equation 
\begin{equation}\label{6.25}
\frac 12\,(M^2-D^2)\,\chi_n=(N_{\mathbb Z}+1/2)\,\chi_n\,.
\end{equation}
Equations \eqref{6.24} and \eqref{6.25} are valid for all $n=0,1,2,\dots$. 
 Again, these operators are closable and unbound.

\section{On the Continuity of the Relevant Operators}\label{continuity}

Let us go back to the sequence of functions $\{\c_n(\phi)\}$. These functions are linearly independent after Proposition A2  (Appendix  \ref{apendb}). Thus, the functions in the sequence $\{\c_n(\phi)\}$ are linearly independent. Then, we may consider the linear space $\mathfrak G$ spanned by them and introduce on it a new scalar product defined as $\langle \c_n|\c_m\rangle=\delta_{nm}$, where $\delta_{nm}$ is the Kronecker delta. This scalar product is now extended to the whole $\mathfrak G$ by linearity to the right and anti-linearity to the left. The resulting pre-Hilbert space may then be completed so as to obtain a Hilbert space that we shall denote as $\mathcal H$. Next, let us define the space $\mathfrak S$ of all functions 
\be\label{7.01}
\ds f(\phi)=\sum_{n=0}^\infty a_n\,\c_n(\phi)\in\mathcal H\,, \qquad a_n\in \C
\ee
 with the following property
\begin{equation}\label{7.1}
\big|\big| f(\phi)\big|\big|^2_p:= \sum_{n=0}^\infty \big| a_n\big|^2\,(n+1)^{2p}<\infty\,,\qquad p=0,1,2,\dots\,.
\end{equation}  
Clearly, $\mathfrak S$ is isomorphic algebraic and topologically to the Schwartz space $\mathcal S$ of all $C^\infty(\mathbb R)$ functions that go to zero at infinity faster than the inverse of any polynomial, see \cite{RSI}, as the topology on $\mathfrak S$ is given by the countable set of norms \eqref{7.1}. The triplet $\mathfrak S \subset \mathcal H \subset \mathfrak S^\times$, where $\mathfrak S^\times$ is the dual space of $\mathfrak S$ endowed with the weak topology corresponding to the dual pair $\{\mathfrak S,\mathfrak S^\times\}$ \cite{HOR} is a rigged Hilbert space or Gelfand triplet. 

The operators $\Phi$, $D_\phi$, and $A_\phi^\pm$ are continuous linear operators on $\mathfrak S$ and the same property holds for the algebra spanned by these operators. For instance, let us pick $f(\phi)\in\mathfrak S$ \eqref{7.01}. Then, using \eqref{6.4}, we have
\be\begin{array} {lll}\label{7.2}
\big|\big|\Phi\,f(\phi)\big|\big|_p &=&\ds 
\frac 12\,\bigg|\bigg|\sum_{n=0}^\infty a_n\,\c_{n+1}(\phi)\bigg|\bigg|_p +
\bigg|\bigg|\sum_{n=0}^\infty a_n\,n\,\c_{n-1}(\phi)
\bigg|\bigg|_p 
\\[0.5cm]\
  & \le & \ds \frac 12 \sqrt{\sum_{n=0}^\infty |a_n|^2\,(n+1)^{2p}} + \sqrt{\sum_{n=1}^\infty |a_n|^2\,n\,(n+1)^{2p}} 
\\[0.6cm]\ds
&  \le & \ds \frac 12 \sqrt{\sum_{n=0}^\infty |a_n|^2\,(n+1)^{2p}} + \sqrt{\sum_{n=0}^\infty |a_n|^2\, (n+1)^{2(p+1)}}  
\\[0.6cm]\ds
  & = &\ds  \frac 12\,\big|\big| f(\phi)\big|\big|_p + \big|\big| f(\phi)|\big|\big|_{p+1}\,.
\end{array}\ee
This proves that $\Phi\,\mathfrak S\subset \mathfrak S$ with continuity. From here, it is obvious that the same property holds for $\Phi^m$, with $m=0,1,2,\dots$. 
The proof for the same property concerning $D_\phi$ comes from \eqref{6.8} and for $A_\phi^\pm$ from \eqref{6.9}. 

From these results, it is clear that all the operators in the algebra spanned by $\Phi$, $D_\phi$ and $A^\pm_\phi$ are continuous on $\mathfrak S$, including the number operator $N_\phi$. 
Then, observe that $A^\pm_\phi$ are the formal adjoint of each other.  From \eqref{6.19}, we see that $D_\phi$ is formally symmetric and that the formal adjoint of $\Phi$ is $-\Phi$.  
Let $B$ be an arbitrary densely defined operator on the Hilbert space and $B^\dagger$ its adjoint. Assume that $B^\dagger$ leaves $\mathfrak S$ invariant, which means that $B^\dagger f\in\mathfrak S$ for any $f\in \mathfrak S$. Then,  using the duality formula
\begin{equation}\label{7.3}
\langle B^\dagger f|G\rangle = \langle f|B G\rangle\,,\qquad f\in\mathfrak S\,,\quad G\in\mathfrak S^\times\,,
\end{equation}
one shows that  $B$ may be extended as a linear operator to the dual $\mathfrak S^\times$. In addition, if $B^\dagger$ is continuous on $\mathfrak S$, so is $B$ on $\mathfrak S^\times$ with the weak topology on $\mathfrak S^\times$. Therefore, the algebra spanned by the operators $\Phi$, $D_\phi$, and $A^\pm_\phi$ may be extended to the dual $\mathfrak S^\times$ and these extensions are continuous with the weak topology on the dual. 

However, the above discussion, notwithstanding its simplicity, relies on unnatural Hilbert metrics and is somehow artificial. On the other hand,   the use of the natural scalar product on $L^2(\mathcal C)$ may require of the introduction of a more artificial test space  that we shall denote with $\mathfrak D$. In order to construct $\mathfrak D$, let us consider the linear space of all {finite} linear combinations of functions $\mathfrak C_n(\phi)$ 
\be\label{7.05}
\sum_{n=0}^N f_n\,\mathfrak C_n(\phi)\,, \qquad f_n\in \C\,,\quad N=0,1,2,\dots. 
\ee
 On this linear space, we define the following set of seminorms  (indeed norms), $\mathfrak p_k(-)$
\begin{equation}\label{7.4}
\mathfrak p_k\left( \sum_{n=0}^N f_n\,\mathfrak C_n(\phi) \right) := \sum_{n=0}^N |f_n| \,b_n\,\left((n+1)!\right)^k\,,\qquad b_n:= e^\pi\,(2\pi)^{2n}\,.
\end{equation}
The resulting locally convex space needs not be complete, although it is always possible to complete it with respect to the locally convex topology generated by the semi-norms \eqref{7.4}. We call $\mathfrak D$ to this completion. Note that
\begin{eqnarray}\label{7.5}
\left|\left| \sum_{n=0}^N f_n\,\mathfrak C_n(\phi) \right|\right| \le \sum_{n=0}^N |f_n|\,||\mathfrak C_n(\phi)|| = \sum_{n=0}^N |f_n|\, \sqrt{\sum_{m=-\infty}^\infty |\psi_n(m)|^2}\,,
\end{eqnarray}
where the identity in \eqref{7.5} is a consequence of \eqref{8.30} and \eqref{8.31}. Then, combining inequalities $\sum_{m=-\infty}^\infty e^{-m^2}\,m^{2n} \le c(n+1)!\,2^{n+1}$ first and then \eqref{8.20} , we obtain that the last term in \eqref{7.5} is smaller or equal to
\begin{equation}\begin{array}{lll}\label{7.6}
\ds \sum_{n=0}^N |f_n|\, 2^n\,(n+1)!\,\sqrt{c\,(n+1)}& \le &\ds   \sqrt{c}\, \sum_{n=0}^N |f_n|\, b_n\,\left((n+1)!\right)^2 \\[0.4cm]
&&\ds  = \sqrt{c}\, \mathfrak p_2\left( \sum_{n=0}^N f_n\,\mathfrak C_n(\phi) \right)\,.
\end{array}\end{equation}
This chain of inequalities shows that the canonical identity $i:\mathfrak D \longmapsto L^2(\mathcal C)$ is continuous, so that
\begin{equation}\label{7.7}
\mathfrak D \subset L^2(\mathcal C) \subset \mathfrak D^\times\,,
\end{equation}
is a rigged Hilbert space, or Gelfand triplet. The dual space $\mathfrak D^\times$ is endowed with any topology compatible with duality (strong, weak, McKey). 

Now, proving the continuity of the operators defined in Sections \ref{sec6.1} and \ref{sec6.2} on $\mathfrak D$ is rather trivial. For instance, take $A^+_\phi$ as defined in \eqref{6.9}. Obviously,
\be\begin{array}{lll}\label{7.8}
\ds A^+_\phi \sum_{n=0}^N f_n\,\mathfrak C_{n}(\phi) &=& \ds\sum_{n=0}^N f_n\, \sqrt{n+1}\, \mathfrak C_{n+1}(\phi) \\[0.4cm] 
&=&\ds  f_0 \,\mathfrak C_1(\phi) + f_1 \sqrt{2}\,\mathfrak C_2(\phi) + \dots + f_n \,\sqrt{n+1}\, \mathfrak C_{n+1}(\phi)\,.
\end{array}\ee
Hence,
\begin{eqnarray}\label{7.9}
\mathfrak p_k\left( A^+_\phi \sum_{n=0}^N f_n\,\mathfrak C_n(\phi) \right)  = |f_0|\, b_1\,\sqrt 1\, [(1+1)]^k + \dots + |f_n|\,b_{n+1}\,\sqrt{n+1}\,\left((n+1)\right)^k\,.
\end{eqnarray}
Then, 
\begin{equation}\label{7.10}
b_{n+1} = e^\pi\,(2\pi)^{2n+2} =b_n (2\pi)^2\,,\qquad \left((n+1)!\right)^k = \left(n+1\right)^k \left(n!\right)^k \le \left(n!\right)^{k+1}\,,
\end{equation}
so that
\begin{equation}\label{7.11}
\mathfrak p_k\left( A^+_\phi \sum_{n=0}^N f_n\,\mathfrak C_n(\phi) \right) \le (2\pi)^2\;  
\mathfrak p_{k+2} \left( \sum_{n=0}^N f_n\,\mathfrak C_n(\phi) \right)\,,
\end{equation}
which proves, both, that $A^+_\phi \mathfrak D\subset \mathfrak D$ and that $A^+_\phi$ are continuous on $\mathfrak D$. Similar proofs apply to the other operators in \eqref{6.1} and \eqref{6.2}.

Analogous results can be obtained when dealing with the operators defined in \mbox{Section~\ref{sec6.3}}. 
\section{Concluding Remarks}

We investigated the role of Hermite functions in Harmonic analysis in connection with Fourier analysis. We showed that Hermite functions permit the construction of  a complete set of periodic functions defined in the unit circle that span $L^2(\mathcal C)$. Using the Gramm--Schmidt procedure, we readily obtain an orthonormal basis for $L^2(\mathcal C)$ out of these~functions. 

At the same time, and using the normalized Hermite functions, we constructed a system of generators in $l_2(\mathbb Z)$, the space of square summable complex sequences indexed by the 
integer numbers. We showed that the use of Fourier series and Fourier transform relates both systems of generators,  in $L^2(\mathcal C)$ and $l_2(\mathbb Z)$ in a very natural way, defining a unitary transformation between these two spaces. 

On the subspace of $L^2(\mathcal C)$, including a complete set of periodic functions, we defined a multiplication and a derivation operator that preserve periodicity in both cases. These operators generate creation and annihilation operators for the defined complete set of periodic functions, which behave just as creation and annihilation operators for the harmonic oscillator. Similar operators with identical properties are defined for $l_2(\mathbb Z)$. We have constructed rigged Hilbert spaces supporting these operators on which they are continuous operators. 

\vspace{+12pt}

\appendix


\section{Determinants of Hermite Polynomials} \label{apendb}

In Appendix  \ref{apendb}, we shall prove some results relevant  for the development of Section \ref{discretefourier}. \medskip

{\bf Proposition A1.-} {sl
The following two determinants are different from zero
\begin{equation}\label{8.15}
\left|\begin{array}{ccccc} H_0(-m) & \dots & H_0(0) & \dots & H_0(m) \\[2ex]
H_1(-m) & \dots & H_1(0) & \dots & H_1(m)   \\[2ex]
\dots & \dots & \dots & \dots & \dots  \\[2ex]
H_{2m}(-m) & \dots & H_{2m}(0) & \dots & H_{2m}(m) \end{array} \right| 
\,,\;
\left| \begin{array}{cccc} H_0(0) & H_0(1) & \dots & H_0(m) \\[2ex] H_1(0) & H_1(1)  &\dots & H_1(m)
\\[2ex] \dots & \dots & \dots & \dots \\[2ex] H_{m}(0) & H_{m}(1) & \dots & H_{m}(m)
 \end{array} \right| \,.
\end{equation}
where $H_n(k)$ is the Hermite polynomial $H_n(x)$ evaluated at the integer point $k$ and  $m\in \N$.
\}\medskip

{\bf Proof.-}

 Let us prove that the second determinant is different from zero. The proof of this property for the first determinant is similar.

We proceed by induction on the dimension $m$ of the determinant. Since we know explicit expressions for the low order Hermite polynomials, we easily conclude that the determinant on the right hand side of \eqref{8.15} is different from zero for $m=0,1,2,3, 4$. Let us assume that this property is true for $m=0,1,2,\dots,n$. Take $m=n+1$ and assume that the property is not correct, i.e., that the determinant vanishes. Then, it must be a linear relation between the rows of the type (let us use column notation)
\begin{equation}\begin{array}{lll}\label{8.16}
\left( \begin{array}{c}H_{n+1}(0) \\ H_{n+1}(1) \\ \vdots \\H_{n+1}(n) \\H_{n+1}(n+1) \end{array}\right) &=& \lambda_0 \left( \begin{array}{c}H_{n}(0) \\ H_{n}(1) \\ \vdots \\H_{n}(n) \\H_{n}(n+1) \end{array}\right) + \lambda_1 \left( \begin{array}{c}H_{n-1}(0) \\ H_{n-1}(1) \\ \vdots \\H_{n-1}(n) \\H_{n-1}(n+1)+ \end{array}\right) +\\[0.4cm]
&& \qquad  \dots +\lambda_n \left( \begin{array}{c}H_{0}(0) \\ H_{0}(1) \\ \vdots \\H_{0}(n) \\H_{0}(n+1) \end{array}\right)\,.
\end{array}\end{equation} 
Those vectors on the right hand side of \eqref{8.16} are linearly independent, since by the hypothesis of induction the vectors formed by their first $n$ components are linearly independent. Therefore, the set of coefficients $\lambda_0,\lambda_2,\dots,\lambda_n$ must be uniquely determined. 

Now, let us consider the relation between the Hermite polynomials given by
\begin{equation}\label{8.17}
H_{n+1}(x)= 2x\,H_n(x) -2n\, H_{n-1}(x)\,,
\end{equation}
and let us use this relation in the last row of \eqref{8.16}, so that
\begin{equation}\label{8.18}
H_{n+1}(n+1) = 2(n+1)\,H_n(n+1) -2n \,H_{n-1}(n+1)\,, 
\end{equation}
which implies that $\lambda_0= 2(n+1)$, $\lambda_1=-2n$ and $\lambda_2 = \dots =\lambda_n=0$. Let us use again \eqref{8.17} in \eqref{8.18}. Hence,  
\begin{equation}\label{8.19}
H_{n+1}(n+1) = [4(n+1)^2 -2n]\, H_{n-1}(n+1)-2(n-1)\,H_{n-2}(n+1)\,,
\end{equation}
so that $\lambda_0=0$, $\lambda_1=4(n+1)^2 -2n$ and $\lambda_2=-2(n-1)$, which is an obvious contradiction. Thus, a relation like \eqref{8.16} is not possible and, therefore, the second determinant in \eqref{8.15} cannot vanish.

We could also have completed the proof by observing the first component. If $n$ is even, then, $H_{n+1}(0)=0$ and $H_{n-1}(0)=0$. Since $\lambda_2=\dots=\lambda_0=0$, this implies that $\lambda_0=0$, a contradiction. If $n$ were odd, an easy calculation shows that $\lambda_1=n$, which is again contradictory with the result obtained with the last row in \eqref{8.16}.
\hfill $\square$\medskip

{\bf Proposition A2.-} {\sl
The sequences  $\chi_n= \{\psi_n(m)\}_{m\in\Z}$, with $\psi_n(x)$ the Hermite functions \eqref{2.1}, for any $n\in \N$ are in $l_2(\mathbb Z)$. They   are linearly independent and span $l_2(\mathbb Z)$.
}\medskip

{\bf Proof.-}
We need to show that, for each value of $n$, $\chi_n:=\sum
\psi_n(m)\,\mathcal B_m$ is a vector in  $l_2(\mathbb Z)$. By the properties of Hilbert spaces, this is equivalent to saying that
\begin{equation}\label{8.20}
\sum_{m=-\infty}^\infty |\psi_n(m)|^2<\infty\,.
\end{equation}
Following the machinery and arguments in the proof of Proposition 1 (Sec.~\ref{periodicfunctions}), we find that
\begin{equation}\label{8.21}
\sum_{m=-\infty}^\infty |\psi_n(m)|^2 \le  2^n\,(n+1)\, (n+1)!\, \sum_{m=-\infty}^\infty e^{-m^2}\,m^{2n}\,.
\end{equation}
The last series in \eqref{8.21} is convergent for any value of $n$. Therefore, the sequences $\chi_n$ are in the Hilbert space $l_2(\mathbb Z)$ for all $n=0,1,2,\dots$. 

To prove that the sequences are linearly independent, let us consider an arbitrary finite linear combination  
 (by a standard definition in linear algebra, all linear combinations are finite) of the form
 \begin{equation}\label{8.22}
0=\sum_{n=0}^N a_n\, \chi_n = \sum_{m=-\infty}^\infty \left\{\sum_{n=0}^N a_n\,\psi_n(m)\right\}\,\mathcal B_m\,.
\end{equation}
Since  $\{\mathcal B_m\}_{m\in\mathbb Z}$ is an orthonormal basis in $l_2(\mathbb Z)$  the change in the order in \eqref{8.22}  is legitimate as well as that the 
\begin{equation}\label{8.23}
\sum_{n=0}^p a_n\,\psi_n(m)=0\,,
\end{equation}
expression that must be valid for $m=0,\pm 1,\pm 2,\dots$. If we choose $m=0,1,2,\dots,n$, \eqref{8.23} gives a linear homogeneous system of $p+1$ equations with $p+1$ indeterminates, which are $a_0,a_1,\dots,a_p$. The determinant of the coefficients of this system is given by
\begin{equation}\label{8.24}
\left| \begin{array}{cccc}  \psi_0(0) & \psi_0(1) & \dots & \psi_0(p)  \\[2ex] \psi_1(0) &  \psi_1(1) & \dots & \psi_1(p) 
\\[2ex] \dots & \dots & \dots& \dots \\[2ex] \psi_p(0) &  \psi_p(1) & \dots & \psi_p(p) 
\end{array}   \right|
= \left| \begin{array}{cccc}  H_0(0) & e^{1/2}\,H_0(1) & \dots & e^{p^2/2}\,H_0(p)  \\[2ex]  H_1(0) &e^{1/2}\,H_1(1) & \dots & e^{p^2/2}\,H_1(p) 
\\[2ex] \dots & \dots & \dots& \dots \\[2ex] H_p(0) & e^{1/2}\,H_p(1) & \dots & e^{p^2/2}\,H_p(p) 
\end{array}   \right|  \,.
\end{equation}
This determinant is different from zero if and only if the second determinant in \eqref{8.15} is non-vanishing, which is true due to Proposition A1. Therefore, we have that $a_0=a_1=\dots =a_p=0$ and the $\{\chi_n\}_{n\in \N}$ are linearly independent. 

Next, let us show that the set of sequences $\{\chi_n\}_{n\in \N}$ spans $l_2(\mathbb Z)$. We know that this happens if and only if for all $n\in \mathbb N$, we have \cite{BN}
\begin{equation}\label{8.25}
\langle \chi_n|\varphi\rangle =0\quad  \Longrightarrow \quad \varphi=\mathbf 0\,, 
\end{equation}
for any $\varphi$ in a dense subspace of $l_2(\mathbb Z)$. 
As a dense subspace, we choose the space of all finite linear combinations of elements of the basis $\{\mathcal B_m\}$. These vectors are of the form
\begin{equation}\label{8.26}
\varphi_p:=\sum_{m=-p}^p a_m\,\mathcal B_m\,,
\end{equation}
where $p$ is an arbitrary non-negative integer and the coefficients $a_m$ are arbitrary.  Let us assume that, for all $\chi_n$, the scalar product $\langle \chi_n|\varphi_p\rangle=0$. We have that
\begin{equation}\label{8.27}
0=\langle \chi_n|\varphi_p\rangle = \sum_{m=-p}^p a_m \langle \chi_n |\mathcal B_m\rangle =  \sum_{m=-p}^p a_m \, \psi_n(m)\,,\qquad \forall n\in\N\,.
\end{equation}
This gives an infinite sequence of algebraic equations with $2 p+1$ indeterminates  $a_m$, with   $m=-p,-p+1\dots,p$. The first $2p+1$   equations of this infinite system are
\begin{equation}\label{8.28}
\begin{array}{ccc}
a_{-p}\,\psi_0(-p)& +\,\dots\dots &+\; a_p\,\psi_0(p)=0\,,\\[2ex]
a_{-p}\,\psi_1(-p)& +\,\dots\dots &+ \, a_p\; \psi_1(p)=0\,,\\[2ex]
\dots\dots & \dots\dots  & \dots\dots  \\[2ex]
a_{-p}\,\psi_{2p}(-p)&+\,\dots\dots &+ \,a_N\;\psi_{2p}(p)=0\,.
\end{array}
\end{equation}
Since the values of the Hermite functions, $\psi_m(x)$, at the integers are given data, this specified set of linear equations with constant coefficients will have a solution different from zero if and only if the determinant  of the coefficients vanishes. This determinant is given in our case by
\begin{equation}\begin{array}{lll}\label{8.29}
\left| \begin{array}{ccccc}
\psi_0(-p) & \dots & \psi_0(0) & \dots & \psi_0(p)\\[2ex]
\psi_1(-p) & \dots & \psi_1(0) & \dots & \psi_1(p)\\[2ex]
\dots & \dots & \dots & \dots & \dots\\[2ex]
\psi_{2p}(-p) & \dots & \psi_{2p}(0) & \dots & \psi_{2p}(p)
\end{array}   \right| \\[0.7cm]
\qquad =
\left|\begin{array}{ccccc}  e^{-\frac{p^2}{2}} H_0(-p) & \dots & H_0(0) & \dots & e^{\frac{p^2}{2}} ,H_0(p)  \\[2ex]
 e^{-\frac{p^2}{2}} H_1(-p) & \dots & H_1(0) & \dots & e^{\frac{p^2}{2}} H_1(p)  \\[2ex]
 \dots & \dots & \dots & \dots & \dots     \\[2ex]
e^{-\frac{p^2}{2}} H_{2p}(-p) & \dots & H_{2p}(0) & \dots & e^{\frac{p^2}{2}} H_{2p}(p)  
\end{array} \right|\,.
\end{array}\end{equation}
This second determinant  is just the determinant in \eqref{8.15} as all the exponentials  cancel out. Then, and due to Proposition A1, the determinant is non-vanishing and consequently, the indeterminates $a_{-p}= \dots = a_0 =\dots =a_p=0$, so that $\varphi_n\equiv\mathbf 0$ and the set 
$\{\chi_n\}_{n\in \N}$ is complete in $l_2(\mathbb Z)$. 
\hfill$\square$

 \section{ Orthonormal Systems in $L^2(\mathcal C)$ and $l_2(\mathbb Z)$} \label{apendc}

In order to construct an orthonormal basis in $L^2(\mathcal C)$ after the set of functions $\{\c_n(\phi)\}$, we first evaluate the scalar product on $L^2(\mathcal C)$ of two of these functions
\be\begin{array}{lll}\label{8.30}
\langle \c_n|\c_m\rangle &=& \ds
\frac{1}{2\pi}\int_{-\pi}^\pi \c_n^*(\phi)\,\c_m(\phi)\,d\phi  \\[0.4cm] 
&=& \ds
 \left(\frac{1}{2\pi}\right)^2\int_{-\pi}^\pi d\phi \sum_{k=-\infty}^\infty \sum_{j=-\infty}^\infty (-i)^n\,i^m \, 
 \psi^*_n(k)\,\psi_m(j)\, e^{-i(-k+j)\phi}  \\[0.4cm] 
&=& \ds
\frac{1}{2\pi}  \sum_{k=-\infty}^\infty \sum_{j=-\infty}^\infty \delta_{k,j}\,i^{m-n}\,\psi_n^*(k)\,\psi_m(j) 
 \\[0.5cm] 
&=& \ds
 \frac{1}{2\pi}\sum_{j=-\infty}^\infty  i^{m-n}\,\psi_n^*(j)\,\psi_m(j) = i^{m-n}\,(\chi_n,\chi_m)\,,
\end{array}\ee
where $(\chi_n,\chi_m)$ is the scalar product in $l_2(\mathbb Z)$ of the sequences $\chi_n$ and $\chi_m$ defined in \eqref{5.6}. 
Due to parity properties for Hermite polynomials,  
the last series in \eqref{8.30} vanishes if one of the indices $n$ or $m$ is even and the other odd. This shows that
\begin{equation}\label{8.31}
\langle \c_n|\c_m\rangle = \left\{ \begin{array}{ccr} 0 & \, & \text{if $n$ is odd and $m$ even or vice versa,}  \\[2ex] \ne 0 \,, & \,& \text{otherwise} \end{array}\right.\,.
\end{equation}
Note that after \eqref{8.30} 
\begin{equation}\label{8.32}
\big|\big|\c_n\big|\big|^2=\langle \c_n|\c_n \rangle=\sum_{m=-\infty}^\infty \big|\psi_n(m)\big|^2\,.
\end{equation}

Now, we are in the position of using the system $\{\c_n(\phi)\}$
to construct an orthonormal system in $L^2(\mathcal C)$. To do it, we need the Gramm--Schmidt process \cite{cheney} (see, for instance,  Ref.~\cite{cheney1} for  a Gramm--Schmidt algorithm in order to use computer techniques). The resulting orthonormal system would be an orthonormal basis for  $L^2(\mathcal C)$ as a consequence of   Proposition A1 (see Appendix  \ref{apendb}). 
As is well known from the set $\{\c_n(\phi)\}$, we can obtain firstly a set of orthogonal functions $\{\d_n(\phi)\}$
where
\be\label{8.33}
 \d_0:= \c_0\,,\qquad 
  \d_{n}:= \c_{n}-\sum_{k=0}^{n-1} \frac{\langle \c_{n}|\d_{k}\rangle}{\langle \d_{k}|\d_{k}\rangle}\,\d_{k}\,.
\ee
However, taking into account  \eqref{8.32}, we can rewrite  the relations \eqref{8.33} as follows distinguishing between  even and odd subindices of the functions, i.e.,
\begin{equation}\begin{array}{lll}\label{8.34}
 \d_0:= \c_0\,,\qquad 
 \ds \d_{2n}&:=& \ds\c_{2n}-\sum_{k=0}^{n-1} \frac{\langle \c_{2n}|\d_{2k}\rangle}{\langle \d_{2k}|\d_{2k}\rangle}\,\d_{2k}\,,\\[0.4cm]
\d_1:= \c_1\,,\qquad 
 \ds \d_{2n+1}&:=&\ds \c_{2n+1}-\sum_{k=0}^{n-1} \frac{\langle \c_{2n+1}|\d_{2k+1}\rangle}{\langle \d_{2k+1}|\d_{2k+1}\rangle}\,\d_{2k+1}\,.
\end{array}\end{equation}
Let us denote as $\cw_n(\phi)$ the normalized functions 
\begin{equation}\label{8.35}
\cw_n(\phi):= \frac{\d_n}{\langle \d_n|\d_n\rangle}\,.
\end{equation}
Due to the properties of the Gramm--Schmidt process, the set of functions $\{\cw_n(\phi)\}_{n\in\N}$ forms an orthonormal system in  $L^2(\mathcal C)$. 
\medskip

Analogously, and using Formulae \eqref{5.12} and \eqref{8.34},  
we may construct new  sequences, $\{\zeta_n\}$, in $l_2(\mathbb Z)$ defined as
\be\label{8.36}
\zeta_0:= \chi_0\,,\qquad 
  \zeta_{n}:= \chi_{n}-\sum_{k=0}^{n-1} \frac{( \chi_{n},\zeta_{k})}{( \zeta_{k},\zeta_{k})}\,\zeta_{k}\,,
\ee
as well as  the normalized series 
\begin{equation}\label{8.37}
\widehat\chi_n:= \frac{\zeta_n}{\langle \zeta_n|\zeta_n\rangle}\,,\qquad n\in \N\,.
\end{equation}
Series \eqref{8.37} gives an orthonormal system in $l_2(\mathbb Z)$. We denote the components of the $\widehat\chi_n$ as $\widehat\chi_n=\{\widehat\psi_n(m)\}_{m\in\mathbb Z}$. After relations \eqref{5.12}, \eqref{8.34} and \eqref{8.36}, it is easy to show that \eqref{5.10} and \eqref{5.11} yield, respectively,
\begin{equation}\label{8.38}
\widehat \c_n(\phi)=\frac{i^n}{\sqrt{2\pi }}  \sum_{m=-\infty}^\infty \widehat \psi_n(m) \,e^{-im\phi}\,,
\end{equation}
and
\begin{equation}\label{8.39}
\widehat\psi_n(m)= \frac{(-i)^n}{\sqrt{2\pi}} \int_{-\pi}^\pi \widehat \c_n(\phi)\,e^{im\phi}\,d\phi\,.
\end{equation}
Thus, we have a  partial isometry $U$ between $L^2(\mathcal C)$ and $l_2(\mathbb Z)$ given for any $f(\phi)\in L^2(\mathcal C)$ in the closed subspace spanned by the $\{\widehat \c_n(\phi)\}$  by
\begin{eqnarray}\label{8.40}
f(\phi)=\sum_{n=0}^\infty \widehat f_n\,\widehat \c_n(\phi)\,; \qquad \big|\big|\widehat c_n\big|\big|^2= \sum_{n=0}^\infty \big|\widehat f_n\big|^2 \,, \qquad  Uf(\phi)= \sum_{n=0}^\infty f_n\,\widehat \chi_n\,.
\end{eqnarray}
After Proposition A1 (Appendix  \ref{apendb}), we conclude that $U$ is a unitary mapping between $L^2(\mathcal C)$ and $l_2(\mathbb Z)$.



\end{document}